\documentclass[aps,twocolumn,showpacs]{revtex4}

\usepackage{graphicx}

\begin{document}

\title{In-flight dissipation as a mechanism to suppress Fermi acceleration.}

\author{Diego F. M. Oliveira}
\author{Marko Robnik}

\affiliation{CAMTP - Center For Applied Mathematics and Theoretical Physics  University of Maribor - Krekova 2 - SI-2000 - Maribor - Slovenia.}

\date{\today} \widetext


\begin{abstract}
Some dynamical properties of time-dependent driven elliptical-shaped billiard are studied. 
It was shown that for the conservative time-dependent dynamics 
the model exhibits the Fermi acceleration [Phys. Rev. Lett. 100, 014103 (2008)]. On the other hand, it was observed that 
damping coefficients upon collisions suppress such phenomenon [Phys. Rev. Lett. 104, 224101 (2010)]. Here, we consider a dissipative model 
under the presence of in-flight dissipation due to a drag force which is 
assumed to be proportional to the square of the particle's velocity. Our results reinforce that dissipation leads to
a phase transition from unlimited to limited energy growth. The behaviour of the average velocity is described using scaling arguments.
\end{abstract}

\maketitle


Dissipative systems have attracted much attention during the last years since they can be used 
 in order to explain different physical phenomena
in different fields of science including atomic and molecular physics
\cite{ref1,ref3}, turbulent and fluid dynamics
\cite{ref4,ref5,ref6}, optics \cite{ref9,ref10},
nanotechnology \cite{ref11,ref12}, quantum and relativistic systems
\cite{ref13,ref15}. Different procedures can be used to
describe such systems. The billiard models are often considered since they can be easily described mathematically
and can be realized experimentally in many different ways, for example, the microwave resonators initiated by H.-J. St\"ockmann in 1990
\cite{refhj} and also superconducting microwave resonators \cite{ref16}, quantum dots \cite{ref17}, 
ultracold atoms \cite{ref18} and many others. From the mathematical point of
view, a billiard is defined by a connected region $Q\subset R^D$, with
boundary $\partial Q\subset R^{D-1}$ which separates $Q$ from its
complement. If $\partial Q=\partial Q(t)$ the system has a
time-dependent boundary and it can exchange energy with the particle
upon collision. In such a case, it is possible to investigate the phenomenon 
called Fermi acceleration, i.e., the unlimited energy growth \cite{ref19}.
According to Loskutov-Ryabov-Akinshin (LRA) conjecture \cite{ref20}, a chaotic component in the phase space with
static boundary is a sufficient condition to observe Fermi acceleration
when a time dependent perturbation is introduced. Results that corroborate the validity
of this conjecture include the time dependent oval billiard
\cite{ref21,ref21a}, stadium billiard \cite{ref22}, Lorentz Gas \cite{ref23}.  Recently, it was shown even that a specific perturbation
in the boundary of an elliptical billiard (integrable) leads to the unlimited energy growth \cite{ref24}. The separatix gives place 
to a chaotic layer and the particles can now experience unlimited energy growth while diffusing in the chaotic layer.

\begin{figure}[t]
\centerline{\includegraphics[width=0.90\linewidth]{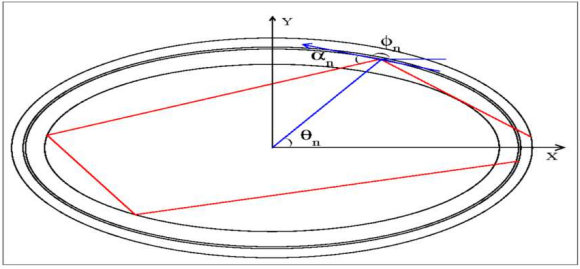}}
\caption{\it{Illustration of five collision with the time dependent boundary. 
The corresponding angles that describe the dynamics are also
illustrated.}}
\label{fig1}
\end{figure}

In this Letter, we will consider a dissipative elliptical billiard with a periodically moving boundary which has been studied in the
pioneering paper in 1996 \cite{ref24b}. Firstly, we assume that the particles 
of mass {\it m} are immersed in a fluid. The dissipative drag force is considered to be proportional to the square of the velocity of 
the particle, $\overrightarrow{V}$. 
To obtain the equation that describes the velocity of the particle along its trajectory, we need to solve Newton's equation where 
$md \overrightarrow{V}/dt=-\eta^\prime \overrightarrow{V}^2$ with the initial velocity $\overrightarrow{V}_n>0$ and
$\eta^\prime$ is the coefficient of the drag force. 
After we introduce the variables $\eta^\prime/m=\eta$, we obtain the velocity 
of the particle as function of time as $\overrightarrow{V}_p(t)= {\overrightarrow{V}_n \over {1+\mid \overrightarrow{V}_n \mid \eta (t-t_n)}}$. 
 We described the model using a four dimensional and non linear map
$T(\theta_{n},\alpha_{n},\arrowvert \overrightarrow{V}_{n} \arrowvert,t_{n})=(\theta_{n+1},
\alpha_{n+1},\arrowvert \overrightarrow{V}_{n+1} \arrowvert
,t_{n+1})$ where the dynamical variables are, respectively, the angular
position of the particle; the angle that the trajectory of the particle
forms with the tangent line at the position of the collision; the
absolute velocity of the particle; and the instant of the hit with the
boundary. 
Figure \ref{fig1} illustrates the geometry of five successive
collisions of the particle with the time-dependent boundary. To obtain
the map, we start with an initial condition
$(\theta_n,\alpha_n,\arrowvert \overrightarrow{V}_n \arrowvert,t_n)$. The Cartesian components of
the boundary at the angular position $(\theta_n,t_n)$ are
\begin{eqnarray}
X(\theta_{n},t_n)&=&[A_0+C\sin(t_n)]\cos(\theta_{n})~, \\ 
Y(\theta_{n},t_n)&=&[B_0+C\sin(t_n)]\sin(\theta_{n})~,
\label{eq2}
\end{eqnarray}
where $A_0$ and $B_0$ are constants, thus, at any time $t_n$ we have elliptical shape. The control parameter $0<C<min(A_0,B_0)$ controls the 
amplitude of oscillation and $\theta\in[0,2\pi)$ is a
counterclockwise polar angle measured with respect to the positive horizontal
axis. The angle between the tangent of the boundary at the position
$(X(\theta_{n}),Y(\theta_{n}))$ measured with respect to the horizontal
line is
$\phi_n=\arctan\left[Y'(\theta_n,t_n) \over X'(\theta_n,t_n) \right]$
where the expressions for both $X'(\theta_n,t_n)=dX(\theta_{n},t_n)/d \theta_n$ and $Y'(\theta_n,t_n)=dY(\theta_{n},t_n)/d \theta_n$.
Since the expressions for $\phi_n$ and
$\alpha_n$ are known, the angle of the trajectory of the particle measured with respect to the positive X-axis is $(\phi_n+\alpha_n)$.
Such information allows us to write the particle's velocity vector as 
$\overrightarrow{V}_n=\vert\overrightarrow{V_p}(t)\vert [\cos(\phi_n+\alpha_n)\widehat{i}+\sin(\phi_n+\alpha_n)\widehat{j}].$
Where $\widehat{i}$ and $\widehat{j}$ denote the unity vectors with respect to the X and Y axis, respectively. 
The particle travels on a straight line until it hits the time dependent boundary.  The position of the particle,
as a function of time, for $t \geq t_n$, is 
$X_p(t)=X(\theta_n,t_n)+r(t)\cos(\phi_n + \alpha_n),$
$Y_p(t)=Y(\theta_n,t_n)+r(t)\sin(\phi_n + \alpha_n).$
Where the sub-index $p$ denotes that such coordinates correspond to 
the particle and $r(t)= {{\eta^{-1}\ln[1+V_n \eta(t-t_n)]}}$, which is the displacement of the particle obtained from direct integration 
of $dr(t)/dt=\overrightarrow{V}_p(t)$. The distance of the particle measured with respect to the origin of the 
coordinate system is given by $R_{p}(t)=\sqrt{X^2_{p}(t)+Y^2_{p}(t)}$ and
$\theta_{p}$ at $X_{p}(t), Y_{p}(t)$ is $\theta_{p}=\arctan[Y_{p}(t)/X_{p}(t)]$. Therefore, the angular
position at $(n+1)^{th}$ collision of the particle with the boundary, i.e. $\theta_{n+1}$, is numerically obtained by solving the following
equation $R_{p}(t)=\sqrt{X^2(\theta_{p},t)+Y^2(\theta_{p},t)}$. The time at $(n+1)^{th}$ collision is obtained 
evaluating $t_{n+1}=t=t_n+t_c$, where $t_c$ is the time during the flight. 
To obtain the new velocity we should note that the referential frame of the
boundary is moving. Therefore, at the instant of collision, the following conditions must be obeyed
\begin{eqnarray}
\overrightarrow{V}_{n+1}\cdot\overrightarrow{T}_{n+1}&=&\overrightarrow{V}_{n}\cdot\overrightarrow{T}_{n+1}~,\\
\overrightarrow{V}_{n+1}\cdot\overrightarrow{N}_{n+1}&=&-\overrightarrow{V}_{n}\cdot\overrightarrow{N}_{n+1}+
2\overrightarrow{V}_{b}(t_{n+1})\cdot\overrightarrow {N}_{n+1}~,
\label{eq7}
\end{eqnarray}
where the $\overrightarrow{T}$ and $\overrightarrow{N}$ are the unitary tangent and normal vectors, respectively, 
and the velocity of the boundary 
$\overrightarrow{V}_{b}(t_{n+1})=C\cos(t_{n+1})[ [\cos(\theta_{n+1})\widehat{i}+\sin(\theta_{n+1})\widehat{j}].$
Then we have
\begin{eqnarray}
|\overrightarrow{V}_{n+1}|=\sqrt{(\overrightarrow{V}_{n+1}\cdot\overrightarrow{T}_{n+1}
)^2+(\overrightarrow{V}_{n+1}\cdot\overrightarrow{N}_{n+1})^2}~.
\label{eq012}
\end{eqnarray}

Finally, the angle $\alpha_{n+1}$ is written as
\begin{eqnarray}
\alpha_{n+1}=\arctan
\left[{\overrightarrow{V}_{n+1}\cdot\overrightarrow{N}_{n+1} \over
\overrightarrow{V}_{n+1}\cdot\overrightarrow{T}_{n+1}} \right]~.
\label{eq013}
\end{eqnarray}

\begin{figure}[t]
\centerline{\includegraphics[width=1.0\linewidth]{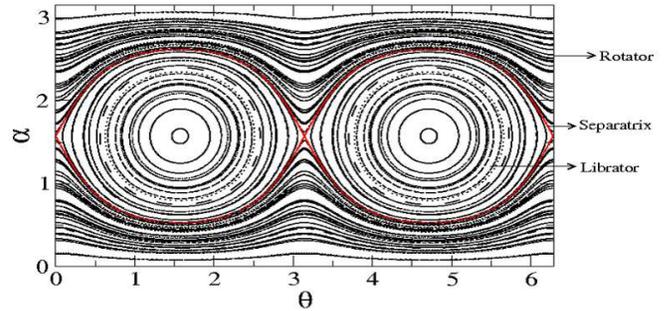}}
\caption{{Phase space for the static elliptical billiard. The control parameters: 
 $A_0=2$ and $B_0=1$.}}
\label{fig3}
\end{figure}

With this four dimensional mapping, we can explore the dynamics of the model.
However, before considering the time-dependent model let us illustrate the
behaviour of the phase space for the static boundary. Indeed, it is well known that the ellipse is an 
integrable billiard system, the product of the two angular momenta with respect to the foci being the integral of motion \cite{refmb,refsinai}.
Figure \ref{fig3}, shows the phase space for
$A_0=2$ and $B_0=1$. We can see a large double island limited by a separatrix and a set of invariant spanning curves. 
We can observe two different kinds of behaviour separated by a separatrix (red curve),
 namely, rotators and librators. Librators consist of trajectories that are confined between the two foci and in the phase 
space are confined by the separatrix curve. On the other hand, rotators are trajectories near to the boundary exploring all the 
values of $\theta$. In the phase space they are outside of the separatrix curve. 

\begin{figure}[t]
\centerline{\includegraphics[width=1.0\linewidth]{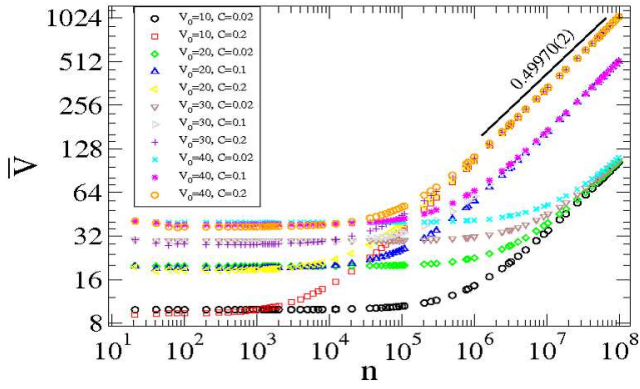}}
\caption{\it{Behaviour of ${\bar V} vs. n$ for different initial
velocities. The control parameters used were $A_0=2$ , $B_0=1$. We have considered the coefficient of the drag force as $\eta=0$.}}
\label{fig19}
\end{figure}

As a part of our numerical results we shall discuss mainly the behaviour of the average velocity of the
particle. Two different procedures were applied in order to obtain the average
velocity. Firstly, we evaluate the average velocity over the orbit for a single initial condition and  then over an ensemble of initial conditions.
Hence, the average velocity is written as
\begin{eqnarray}
\overline{V}={{1}\over{M}}\sum_{i=1}^M{{1}\over{n+1}}\sum_{j=0}^nV_{i,j}~,
\label{eq16}
\end{eqnarray}
where the index $i$ corresponds to a sample of an ensemble of initial conditions, $M$ denotes the number of different initial conditions. 
We have considered $M=200$ in our simulations.
It was shown by Lenz et al. \cite{ref24,ref24a} that when a driving perturbation is introduced into the system, opposite to the expectations, it presents 
 a phenomenon known as Fermi acceleration or unlimited energy gain. Such a behaviour happens because when the driving amplitude $C \ne 0$
the separatix is replaced by a chaotic layer. A particle which starts its dynamics in a rotator orbit can change its dynamics to a librator 
and vice versa. The chaotic diffusion within the chaotic layer leads to unlimited energy growth. Figure \ref{fig19}
shows the behavior of the average velocity as a function of the number of collisions. We have considered the conservative 
case where the drag coefficient is $\eta=0$. For such a case the particle's velocity is 
$\overrightarrow{V}_p(t)=\overrightarrow{V}_n$ and $r(t)= \arrowvert \overrightarrow{V}_n \arrowvert (t-t_n)$. 
As one can see, all curves of the $\bar V$ behave quite similarly in the sense that: (a) for short $n$, the average velocity remains constant
for a while, but eventually, (b) after a crossover, all the curves start growing with the same exponent. This is at variance with the result 
obtained by Lenz et al \cite{ref24}, perhaps because they were not considering the average velocity for large enough values of $n$.

We discuss now the effect of dissipation introduced via frictional force. To obtain the average velocity, we randomly choose $t \in [0, 2\pi]$,
$\theta \in [0, 2\pi]$ and $\alpha \in [0, \pi]$. We also fix the value of $\eta=10^{-3}$. Additionally, in order to avoid the initial 
plateau we also have fixed initial velocity as 
$V_0=10^{-5}$. The model of collisional dissipation by Leonel and Bunimovich \cite{ref27} is different from
our model of in-flight dissipation due to the drag force in detail, but should behave similarly in the statistical sense (on the average), especially in
the chaotic regime, because then we have $<V_{n+1}>=<V_n>e^{-\eta r}$, thus the effective damping coefficient $\delta=e^{-\eta <r>}$, where
$<r>$ is the mean free path of the particle.

\begin{figure}[t]
\centerline{\includegraphics[width=1.0\linewidth]{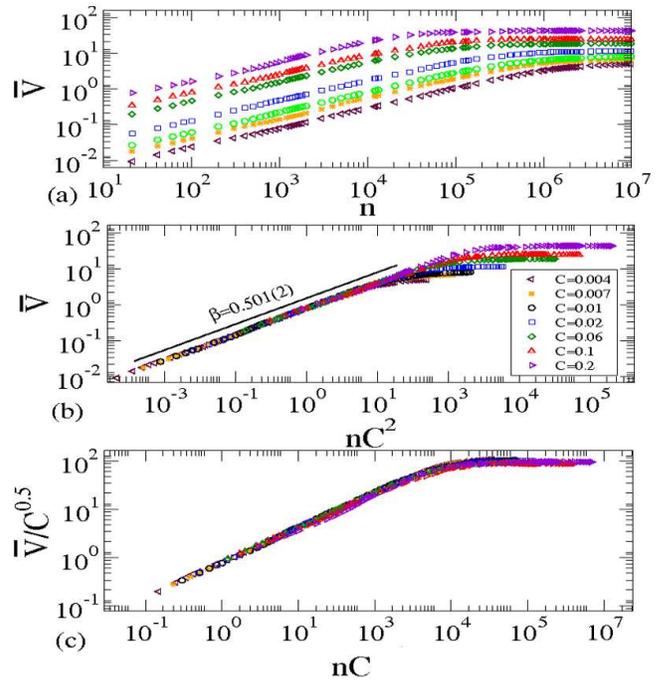}}
\caption{(a) Behaviour of the average velocity as function of $n$ for different values of the control parameter $C$. 
(b) Their initial collapse after the transformation $nC^2$. (c) Their collapse onto a single universal plot.}
\label{fig33}
\end{figure}

In Fig. \ref{fig33} (a) we show the behaviour of the average velocity as a
function of the number of collisions for different values of $C$.
Note that for different values of $C$ and for short $n$, the
average velocity starts to grow and then it bends towards a regime
of saturation for long enough values of $n$. It must be emphasized that
different values of the parameter $C$ generate different
behaviors for short $n$. However, applying the transformation
$n\rightarrow n {C}^2$ coalesces all the curves at short $n$, as
shown in Fig. \ref{fig33}(b). For
such a behaviour, we can also propose the following scaling hypotheses:
(i) When $n\ll{n_x}$ the average velocity is 
\begin{equation}
\overline{V}(nC^2,C)\propto (nC^2)^{\beta}.
\label{eq17}
\end{equation}
where the exponent $\beta$ is called the acceleration exponent.
(ii) When $n\gg{n_x}$, the average velocity is described as
\begin{equation}
\overline{V}_{sat}\propto C^{\gamma}.
\label{eq018}
\end{equation}
where $\gamma$ is the saturation exponent.
(iii) The crossover from growth to the saturation is written as
\begin{equation}
n_x\propto C^{z}~.
\label{eq019}
\end{equation}
where $z$ is called crossover exponent.

\begin{figure}[t]
\centerline{\includegraphics[width=1.0\linewidth]{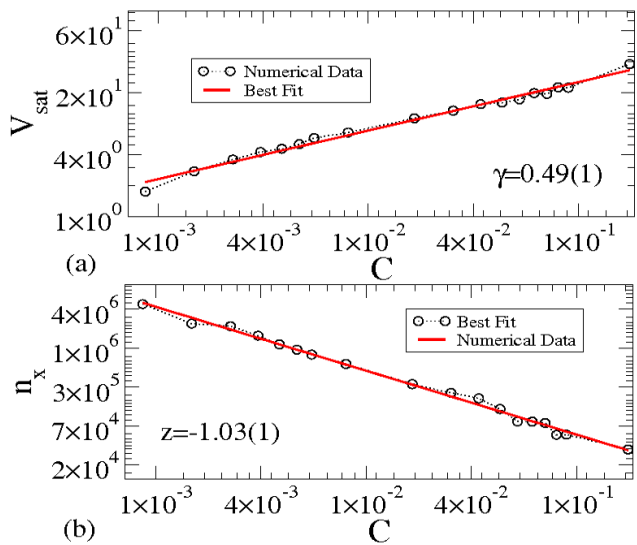}}
\caption{(a) Plot of $\bar V_{sat}$ as function of the control parameter $C$. (b) Behaviour of the crossover number $n_x$ against $C$.}
\label{fig4}
\end{figure}

These scaling hypotheses allow us to describe the average velocity in terms of a scaling function of the type
\begin{eqnarray}
\overline{V}[nC^2,C]=\lambda\overline{V}[{\lambda}^p{nC^2},{\lambda}^q{C}]~,
\label{eq720}
\end{eqnarray}
where $p$ and $q$ are scaling exponents and $\lambda$ is a scaling
factor. Moreover, $p$ and $q$ must be related to $\beta$, $\gamma$ and $z$. Because ${\lambda}$ is a scaling factor, we can specify that
${\lambda}^{p}nC^2=1$, yielding
\begin{eqnarray}
\overline{V}[nC^2,C]={(nC^2)}^{-1/p}\overline{V}_1[n^{-q/p}C]~,
\label{eq721}
\end{eqnarray}
where
$\overline{V}_1[(nC^2)^{-q/p}C]=\overline{V}[1,(nC^2)^{-q/p}C]$ is assumed to be constant for $n\ll{n_x}$. Comparing Eqs.
(\ref{eq17}) and (\ref{eq721}), we obtain $\beta=-1/p$. Choosing now ${\lambda}^q C=1$, we find that ${\lambda}=C^{-1/q}$ 
and Eq. (\ref{eq720}) is given by
\begin{eqnarray}
\overline{V}[nC^2,C]=C^{-1/q}\overline{V}_2[C^{-p/q}nC^2]~,
\label{eq722}
\end{eqnarray}
where $\overline{V}_2[C^{-p/q}nC^2]=\overline{V}[C^{-p/q}nC^2,1]$ is assumed to be constant for $n\gg{n_x}$. 
Comparing Eqs. (\ref{eq018}) and (\ref{eq722}), we obtain
$\gamma=-1/q$ [see Fig. \ref{fig4} (a)]. A power law fitting in Fig. \ref{fig4} gives us that $\beta=0.501(2)$. 
Such value was obtained from the
range of $C\in[10^{-3},2\times10^{-1}]$. Given the two values of the scaling factor {$\lambda$}, one can easily conclude that
$z={\gamma \over \beta}-2=-0.97(1)$, which is in excellent agreement with the value obtained numerically, as shown in Fig.
\ref{fig4} (b). A confirmation of the initial hypotheses is made by a
collapse of all the curves of ${\bar V} vs. n$ onto a single and
universal plot, as shown in Fig. \ref{fig33} (c), showing that the system is
scaling invariant under specific transformation. 
With this good collapse of all the curves of the average velocity and considering that the critical 
exponents are $\beta\cong 0.5$, $\gamma\cong 0.5$ and $z\cong -1$, we can conclude
that the time dependent driven elliptical billiard belongs to the same class of universality of the one 
dimensional Fermi-Ulam model \cite{ref25} and the periodically corrugated waveguide \cite{ref26}.
The scaling can also be described in terms of is the coefficient of the drag force $\eta$. In such a case $V_{sat} \propto {\eta^{-0.521(3) }}$ and 
$n_{x} \propto {\eta^{ -1.070(6) }}$. We have fixed $C=0.1$ and $V_0=10^{-5}$. Therefore, $\eta \rightarrow 0$ implies 
that $V_{sat}$ and $n_{x}$
 both diverge, thus recovering the results for the conservative case, i.e., exhibiting Fermi acceleration.
Additionally, our results reinforce
that in-flight dissipation  is a sufficient condition to suppress the phenomenon of Fermi acceleration like in the case of collisional 
dissipation \cite{ref27}.

As the concluding remark, we have studied some dynamical properties of a time dependent driven elliptical billiard. 
In the static case the phase space is integrable where two kinds of trajectories are observed: rotator and librator.
We have introduced time dependent perturbations and in-flight dissipation. We have observed that average velocity grows for small number of 
collision and then, after a crossover, it reaches a regime of saturation for large $n$. Thus we do not observe the unlimited energy growth 
(Fermi acceleration). We have also studied the behaviour of the average velocity using scaling arguments. 
We have shown that there is a relation between the critical exponents $\gamma$, $\beta$ and $z$. Our scaling hypotheses are 
confirmed by a perfect 
collapse of all the curves onto a single universal plot. Additionally, we confirm that the two dimensional elliptical model belongs to the 
same class of universality of the Fermi-Ulam model (1-D) and the corrugated wave guide (1-D), for the range of control parameters studied.

D.F.M.O gratefully acknowledges Ad futura Foundation - Slovenia for financial support.  M. R. acknowledges the financial support of 
The Slovenian Research Agency.

\end{document}